\documentclass[a4paper,12pt]{article}

\usepackage{amsmath,amssymb,algorithmic}
\usepackage{latexsym}
\usepackage[latin1]{inputenc}
\usepackage{ulem}

\usepackage[pdftex]{graphicx}
\usepackage[frenchb]{babel}
\usepackage{babelbib}
\usepackage{hyperref}
\hypersetup{
colorlinks=true, 
citecolor=blue, %
pdfmenubar=false,
pdftoolbar=false,
bookmarksopen=true,
breaklinks=true, 
urlcolor=blue, 
linkcolor=blue, 
}

\title{Wikistat 2.0: Ressources pédagogiques pour \xout{la Sciences des Données} l'Intelligence Artificielle}

\author{Philippe Besse\thanks{Université de Toulouse -- INSA, Institut de Mathématiques de Toulouse, UMR CNRS}\and Brendan Guillouet\thanks{Université de Toulouse -- INSA, IRT Saint Exupéry} \and Béatrice Laurent\thanks{Université de Toulouse -- INSA, Institut de Mathématiques de Toulouse, UMR CNRS} }

\DeclareGraphicsExtensions{.pdf,.jpg,.png}
\graphicspath{{Figures/}{../}}
\begin{document}
\sloppy

\maketitle

{\bf Résumé}: 
{\it Big data}, science des données, {\it deep learning}, intelligence artificielle, sont les mots clefs de battages médiatiques intenses en lien avec un marché de l'emploi en pleine évolution qui impose d'adapter les contenus de nos formations professionnelles universitaires. Quelle intelligence artificielle est principalement concernée par les offres d'emplois? Quelles sont les méthodologies  et technologies qu'il faut privilégier dans la formation?  Quels objectifs, outils et ressources pédagogiques est-il nécessaire de mettre en place pour répondre à ces besoins pressants?  Nous répondons à ces questions en décrivant les contenus et ressources opérationnels dans la spécialité Mathématiques appliquées, majeure Science des Données, de l'INSA de Toulouse. L'accent est mis sur une formation en Mathématiques (Optimisation, Probabilités, Statistique) fondamentale ou de base associée à la mise en \oe uvre pratique des algorithmes d'apprentissage statistique les plus performants, avec les technologies les plus adaptées et sur des exemples réels.  Compte tenu de la très grande volatilité des technologies, il est impératif de former les étudiants à l'autoformation qui sera leur outil de veille technologique une fois en poste; c'est la raison de la structuration du site pédagogique \href{https://github.com/wikistat/}{\tt github/wikistat} en un ensemble de tutoriels.  Enfin, pour motiver la pratique approfondie de ces tutoriels, un jeu sérieux est organisée chaque année sous la forme d'un concours de prévision entre étudiants de masters de Mathématique Appliquées pour l'IA.

{\bf Mots-clefs}: science des données, intelligence artificielle, apprentissage statistique, données massives, enseignement, jeux sérieux.

\centerline{\Large Wikistat 2.0: Educational Resources for Artificial Intelligence}
\vspace*{2cm}

{\bf Abstract}: 
Big data, data science, deep learning, artificial intelligence are the key words of intense hype related with a job market in full evolution, that impose  to adapt the contents of our university professional trainings. Which artificial intelligence is mostly concerned by the job offers? Which methodologies and technologies should be favored in the training pprograms? Which objectives, tools and educational resources do we needed to put in place to meet these pressing needs? We answer these questions in describing the contents and operational ressources in the Data Science  orientation of  the speciality  Applied Mathematics at INSA Toulouse.  We focus on basic mathematics training (Optimization, Probability, Statistics), associated with the practical implementation of the most performing statistical learning algorithms, with the most appropriate technologies and on real examples. Considering the huge volatility of the technologies, it is imperative to train students in seft-training, this will be their technological watch tool when they will be in professional activity. This  explains  the structuring of the educational site \href{https://github.com/wikistat/}{\tt github/wikistat} into a set of tutorials. Finally, to motivate the thorough practice of these tutorials, a serious game is organized each year in the form of a prediction contest between students of Master degrees in Applied Mathematics for IA.


\newpage
\section{Introduction}
\subsection{Battage médiatique, scientifique et commercial}

{\it Big Data Analytics, Data Science, Machine Learning, Deep Learning}, Intelligence Artificielle, un battage médiatique ({\it buzz word}) en chasse un autre, reflets ou écumes de disruptions technologiques importantes et surtout continuelles. Quels objectifs et choix pédagogiques engager pour anticiper les contenus de nos formations et les modes d'acquisition des compétences afin de préparer efficacement l'intégration des nouveaux diplômés? C'est à ces questions que nous tâchons d'apporter des éléments de réponses, non pas des réponses  théoriques ou des déclarations d'intention, mais plutôt des retours d'expériences et de réalisations en constante (r)évolution au sein de la spécialité Mathématiques Appliquées de l'INSA de Toulouse.

Il est évidemment important de communiquer avec les bons intitulés et les étudiants de l'INSA l'ont bien compris en se déclarant {\it data scientist} sur leur CV depuis 2013. Mais les bons choix d'investissement, ceux pédagogiques qui prennent du temps et engagent sur la durée, ne peuvent être pris en s'attachant à l'écume des mots, même inlassablement soulevée par des rapports officiels, média ou {\it hashtags} des réseaux sociaux. Il ne suffit pas de changer un intitulé de cours ou de diplôme.

Alors que les ressources pédagogiques, les MOOCs, SPOCs, tutoriels, se déversent à profusion sur internet, que devient le rôle d'un enseignant et plus précisément d'un enseignant / chercheur? Certes contribuer à produire de la connaissance par la recherche mais, en responsabilité pédagogique, une fonction essentielle consiste à prioriser des choix. Sous l'écume médiatique, quelles sont les méthodes, les technologies, les algorithmes, dont les performances, donc la diffusion, motivent le temps et l'implication nécessaires à leur intégration dans un cursus académique inexorablement contraint par le volume horaire?

La pression médiatique n'est pas seule en jeu, il faut noter aussi celle, académique, de publication: {\it publish or perish}, qui conduit à la production de milliers d'articles décrivant l"invention" de centaines de méthodes, algorithmes, librairies et de leurs très nombreuses variantes incrémentales, alors qu'en pratique, il faut reconnaître que les différences de performance n'apparaissent pas toujours significatives. Lire à ce sujet les articles de Hand (2006) et Donoho (2015) très critiques envers les algorithmes d'apprentissage récents, et pas seulement pour opacité et manque d'interprétabilité. 

Notons aussi la pression commerciale ou publicitaire des entreprises, {\it start-up} ou grands groupes, disputant les parts d'un marché en forte croissance mais très volatil ou versatile. Chaque année depuis 2012 et motivés par des besoins de visibilité économique, Matt Turck et ses collaborateurs de la société {\it Firstmark} proposent une \href{http://mattturck.com/wp-content/uploads/2018/06/Matt-Turck-FirstMark-Big-Data-Landscape-2018.png}{représentation graphique} du paysage ou de l'écosystème, devenu fort complexe, des entreprises traitant de données massives ou plutôt maintenant de données massives et d'intelligence artificielle. 
Ils tâchent chaque année de prendre en compte les créations, disparitions, fusions, des entreprises du domaine.


\subsection{Quelle Intelligence artificielle?}
Les entreprises ayant appris à stocker, gérer massivement leurs données depuis 10 ans, la phase suivante concerne leur analyse pour leur valorisation et l'aide à la décision, voire de la décision automatique. Après s'être appelée  "{\it big data analytics}" puis "{\it data science}" cette phase fait maintenant référence à une pratique d'{\it intelligence artificielle} (IA), appellation largement médiatisée, notamment depuis les succès remarquables en reconnaissance d'images depuis 2012, en traduction automatique, d'AlphaGo en 2016 ou autour des expérimentations de véhicules autonomes. 

L'IA n'est pas une invention récente car cette discipline ou plutôt cet ensemble de théories et techniques est apparue conjointement avec le développement des tous premiers ordinateurs (ENIAC en 1943), eux-mêmes conséquences des efforts, durant la deuxième guerre mondiale, pour produire rapidement des abaques de balistique puis réaliser les calculs de faisabilité de la première bombe atomique. L'objectif initial était la simulation des comportements du cerveau humain. C'est aussi en 1943 que Mc Culloch (neurophysiologiste) et Pitts (logicien) ont proposé les premières notions de {\it neurone formel}. Notons le début de la théorisation de l'IA avec les travaux pionniers d'Alan Turing en 1950 et la première apparition de la notion de {\it perceptron}, le premier réseau de neurones formels, par Rosenblatt en 1957. Manque de moyens de calcul et d'algorithmes pertinents, l'approche connexionniste de l'IA est mise en veilleuse durant les années 70 au profit de la logique formelle ({\it e.g.} calcul des prédicats du premier ordre) comme outil de simulation du raisonnement. Les {\it systèmes experts} associant base de connaissance (règles logiques), base de faits et moteur d'inférence ont connu un certain succès, notamment avec le développement du langage {\it Prolog}, mais on butté sur la complexité algorithmique explosive des problèmes NP complets. Ce fut alors, au début des années 80, le retour massif de l'approche connexionniste avec le développement de l'algorithme de rétropropagation du gradient qui a ouvert la possibilité, en lien avec des moyens de calculs suffisamment performants, de l'apprentissage de réseaux de neurones complexes. Le développement de l'IA s'est ensuite focalisé dans les années 90 sur des objectifs d'apprentissage ({\it machine learning}), qui devint plus précisément l'apprentissage statistique ({\it statistical learning}) en conséquence de la publication du livre éponyme de Vapnik (1995). C'est toute une farandole d'algorithmes: séparateurs à vaste marge (SVM), {\it bagging, boosting, random forest}... qui provoqua à nouveau la mise en retrait des approches connexionnistes considérées, au même titre que bien d'autres algorithmes souvent plus performants. Mais certains chercheurs dont Yan le Cun, Yoshua Benjio, Geoffrey Hinton, ont continué à développer des structures connexionnistes spécifiques dont les fameux réseaux intégrant des produits de convolution ({\it convolutional neural netwoks})  sur des images. L'accumulation complexe de ces couches fut nommée apprentissage profond ({\it deep learning}) avec un réel succès marketing.

Actuellement, la présentation médiatique de l'IA diverge rapidement vers des questions philosophiques (transhumanisme), comme celle de {\it singularité technologique} lorsque les machines deviendront plus "intelligentes" que l'homme... Le développement de l'IA soulève des questions également anxiogènes de destruction de nombreux emplois qualifiés (Stiegler et Kourou 2015) et pas seulement des métiers manuels absorbés par la robotisation des entreprises. D'autres craintes sont liées aux menaces concernant la vie privée ainsi qu'aux questions éthiques abordées par ailleurs (Besse et al. 2017). Nous oublierons ces aspects pour nous focaliser sur les algorithmes d'IA en exploitation, ceux qui impactent nos quotidiens professionnels ou personnels, conséquences de la {\it datafication} de nos environnements. Ces algorithmes, capables de s'entraîner, en surfant sur la vague ou plutôt le tsunami des données, afin de construire des décisions automatiques, constituent le sous-ensemble historique de l'IA appelé apprentissage automatique ou {\it machine learning}. Plus précisément, nous laisserons de côté les algorithmes de renforcement ou de décision séquentielle ({\it e.g.} bandit manchot) qui sont des algorithmes d'optimisation stochastique trouvant leurs applications dans la gestion des sites de vente en ligne. Il reste alors le principal sous-ensemble des algorithmes d'{\it apprentissage statistique} au sens de Vapnik (1995), incluant également l'apprentissage profond ou {\it deep learning}. Ceux-ci construisent des règles de décision ou des prévisions par minimisation d'un \href{http://wikistat.fr/pdf/st-m-app-risque.pdf}{risque}, généralement une erreur moyenne de prévision. Leur succès et la généralisation de leur utilisation sont des conséquences directes de la datafication ({\it big data}) du quotidien.

Marketing et {\it data mining}, finance et {\it trading} algorithmique, traduction automatique et traitement du langage naturel ({\it sentiment analysis}), reconnaissance faciale et analyse d'images en lien par exemple avec les véhicules autonomes, aide au diagnostic, détection d'anomalies, prévision de défaillance et maintenance préventive dans l'industrie... sont autant de domaines d'application des algorithmes d'apprentissage statistique, sous-ensemble de l'Intelligence Artificielle bénéficiant et  valorisant les masses de données en croissance exponentielle.

\subsection{Quels choix?}
Au sein de cet environnement, ou de cette jungle, une forme de sélection naturelle opère de façon drastique sur les méthodes et technologies associées, celles qui font leurs preuves, s'adaptent et survivent au fil des mises à jour des librairies logicielles et les autres qui s'éteignent car finalement inefficaces ou inadaptées au changement environnemental, par exemple au passage aux échelles volume, variété, vélocité, des données. Trois facteurs ou observatoires semblent déterminants pour suivre cette évolution en temps réel. 
\begin{itemize}
\item Le logiciel libre et les librairies associées, qu'elles soient accessibles en R, Python, Spark... envahissent inexorablement le paysage; {\it Google} ne s'y est pas trompé en ouvrant les codes de \href{https://www.tensorflow.org/}{\it Tensor Flow} et simplifiant son accès avec \href{https://keras.io/}{\it Keras}. Les marges financières significatives ne sont plus apportées par la vente de logiciels mais par celle de services; seuls des logiciels libres autorisent les expérimentations indispensables et permanentes des méthodes et algorithmes dans des situations toujours renouvelées.
\item Corrélativement au logiciel libre, le travail, et pas seulement le développement de codes, devient agile et surtout {\it collaboratif}. L'environnement {\it git} et plus particulièrement l'expansion du site \href{https://github.com/}{\tt github} en sont des révélateurs; {\it Microsoft} ne s'y est pas trompé en rachetant {\it Github}.  Le succès de structures comme celle des Instituts de Recherche Technologiques en sont d'autres exemples qui dépassent les problèmes de propriété industrielle. Les données propriétaires restent confidentielles mais idées, méthodes et même portions de codes sont partagées, co-développées. Suivre à ce sujet le développement du projet franco-québécois DEEL  ({\it dependable and explainable learning}) piloté en France par l'IRT St Exupéry et associant ingénieurs, chercheurs industriels et académiques, 
\item Les suivis et soutenances de stages, de projets industriels, les encadrements de thèses CIFRE, les premières embauches sont autant d'exemples d'expérimentations en vraie grandeur. Les responsabilités, pédagogique d'un diplôme et celle scientifique de projets de recherche, constituent un poste d'observation de premier plan, même si biaisé par la localisation géographique. Cette position permet d'identifier ce qui marche, ou pas, en fonction des domaines d'applications, qu'ils soient industriels ou publics, aéronautiques, médicaux ou autre. 
\end{itemize}

En définitive, le développement des ressources pédagogiques disponibles sur le nouveau site \href{https://github.com/wikistat/}{\tt github.com/wikistat} qui fait suite et vient compléter \href{http://wikistat.fr/}{\tt wikistat.fr}, est la conséquence de ces remarques pilotant un ensemble de choix: 

\begin{itemize}
\item choix technologiques en perpétuelles (r)évolutions ou disruption associés à des 
\item choix et objectifs pédagogiques avec en conséquence des
\item moyens pédagogiques à mettre en \oe uvre.
\end{itemize}

Cette façon schématique de séquencer le propos n'est évidemment pas représentative de la dynamique de la démarche. Elle enchaîne sur l'exposé de Besse et Laurent (2015) qui tentait déjà de définir ou plutôt caractériser la fausse nouvelle {\it Science des Données} avant de détailler l'évolution en cours et à venir du cursus de la spécialité Mathématiques Appliquées. Revenir à cet article, pourtant récent, permet d'identifier des technologies déjà abandonnées, au moins dans les cours, comme {\it Mahaout, RHadoop, H2O...}, d'autres maintenues ou renforcées: {\it Spark}, et certaines annoncées puis effectivement introduites {\it TensorFlow}.

\section{Choix technologiques}
Détaillons les technologies retenues et les quelques raisons qui ont présidé à ces choix.
\subsection{\it Hadoop}
\begin{figure}
\centerline{\includegraphics[width=12cm]{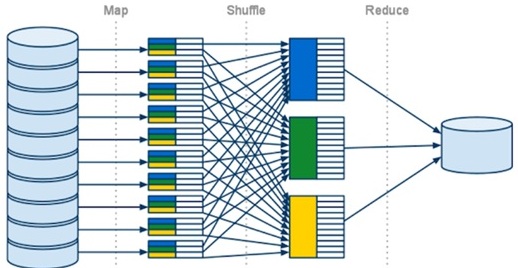}}
\caption{\it Organisation schématique des fonctionnalités  Map-Reduce sous Hadoop Distributed File System (HDFS).}\label{hadoop}
\end{figure}

Une architecture de données distribuées associée à un système {\it Hadoop} de gestion de fichier est devenue la technologie caractéristique voire même emblématique des données massives. Celle-ci offre des capacités de stockage et de parallélisation des traitements incontournables mais pose des contraintes très fortes aux algorithmes susceptibles d'y être exécutés en itérant  nécessairement les seules fonctions (cf. figure \ref{hadoop}): {\it map} parallélisable, {\it schuffle} implicite de répartition vers celle {\it reduce} d'agrégation des résultats. Un traitement efficace de données massives est obtenu à condition de ne pas les déplacer (temps de transfert), ce sont les algorithmes ou codes de calculs qui sont transférés, et de ne les lire qu'une seule fois (temps de lecture disque). C'est ainsi que certains algorithmes survivent à ces contraintes ({\it e.g.} $k$-{\it means}) et sont développés dans les librairies afférentes, tandis que d'autres ({\it e.g.} $k$ plus proches voisins) ne passent pas à cette échelle et prennent la voie de l'extinction.

Plus précisément, la contrainte d'une lecture unique impose de conserver en mémoire les données entre deux itérations d'un algorithme complexe afin d'économiser des accès disques rédhibitoires. C'est justement la principale fonctionnalité qui a fait l'originalité et le succès de {\it Spark} avec le principe de {\it resilient data set}. 

\subsection{\href{https://spark.apache.org/}{\it Spark}}
\begin{figure}
\centerline{\includegraphics[width=12cm]{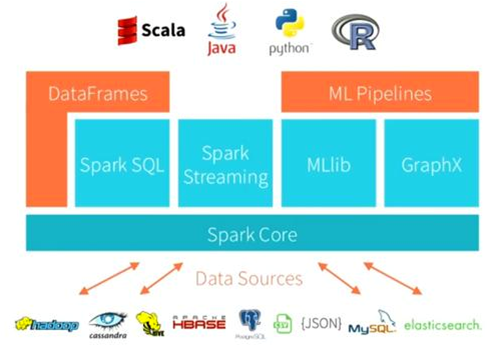}}
\caption{\it La technologie Spark et son écosystème.}\label{spark}
\end{figure}

Cette technologie peut être comprise comme une couche logiciel ({\it framework}) au-dessus d'un système de gestion de fichiers et intégrant une programmation ({\it map/reduce}) fonctionnelle. En plus de gérer des fichiers de données {\it résilients} donc conservés en mémoire entre deux itérations, {\it Spark} propose un ensemble de fonctionnalités (cf. figure \ref{spark}) permettant d'adresser tout système de gestion ou type de fichiers: {\it Hadoop, MongoDB, MySQL, ElasticSearch, Cassandra, HBase, csv, JSON...} ainsi que des capacités de requêtes SQL. \`A cela s'ajoutent des modules qui offrent des possibilités de traiter des données en temps réel ({\it streaming}) ainsi que des graphes. Un dernier module ({\it MLlib}) propose une librairie d'algorithmes d'apprentissage supervisés et non-supervisés adaptés au passage à l'échelle volume. 

Enfin, un des atouts de {\it Spark} et non des moindres, est l'API {\it PySpark} qui permet de coder en Python toutes les opérations citées précédemment et, pour le béotien qui dispose d'une formation plus axée Mathématiques appliquées qu'Informatique, c'est très utile. Il n'a pas à apprendre encore d'autres langages de programmation spécifiques aux données massives ({\it Hive, Pig}) ou généralistes ({\it Java, Scala}) mais au c\oe ur de {\it Spark} pour aborder efficacement ce domaine.

\subsection{Python {\it vs.} R}

La section précédente justifie déjà le choix de Python comme langage permettant de gérer, traiter, analyser des données massives et distribuées. Plus précisément, Python est un langage interprété à tout faire, ce qui présente de nombreux avantages mais aussi des inconvénients. Python est un langage procédural comme C ou Java, c'est aussi, grâce à la librairie {\it numpy}, un langage matriciel comme Matlab ou R. C'est enfin un langage qui intègre des commandes fonctionnelles ({\it e.g. map, reduce}) comme Lisp. Les inconvénients sont de nature pédagogique car un langage à tout faire permet aussi d'écrire des programmes n'importe comment et ce, d'autant plus que beaucoup de déclarations sont implicites. Néanmoins, ceci n'a pas empêché Python d'être choisi comme langage pédagogique dans beaucoup d'établissements et, confronté à des données un peu volumineuses, un étudiant comprend vite ce qu'il faut éviter de programmer avec un langage matriciel et / ou un langage fonctionnel pour obtenir des exécutions efficaces. Enfin  et cette raison peut suffire, Python remplace Matlab dans beaucoup d'environnements industriels, notamment dans l'aéronautique et l'espace, et le système SAS dans bien d'autres domaines.

Si Python présente tant d'avantages, pourquoi conserver l'enseignement et l'utilisation de R dont les capacités de parallélisation, tout du moins avec le noyau actuel, sont plus complexes à mettre en \oe uvre surtout sous Windows. Certes Python se montre généralement plus efficace avec en plus la possibilité de pré-traiter les données en les laissant sur disque lorsqu'elles sont trop volumineuses pour être intégralement chargées en mémoire comme l'impose R. Néanmoins, les fonctionnalités des librairies Python, notamment en Statistique et en aides graphiques à l'interprétation, sont largement inférieures à ce qui est proposé en R. Aussi, pour aborder des questions statistiques classiques, déployer des méthodes d'exploration multidimensionnelles avec les graphes afférents, comprendre et interpréter des modèles ou des arbres de décision, R est beaucoup mieux armé que Python. Ces deux environnements apparaissent finalement comme très complémentaires.

Besse et al. (2016) développent une comparaison détaillée de trois environnements: Python {\it Scikit-learn}, R, {\it Spark MLlib}, pour l'apprentissage sur des cas d'usage de données presque massives: reconnaissance de caractères, système de recommandation de films, traitement de grandes bases de textes. Celle-ci met clairement en évidence la puissance de parallélisation de {\it Spark} pour l'exécution de certaines étapes d'analyse, notamment tout ce qui concerne la préparation des données ou {\it data munging}. En revanche, l'exécution de certains algorithmes plus sophistiqués de la librairie {\it MLlib} provoque des dépassements mémoires et donc des plantages intempestifs. En effet, l'apprentissage de données massives peut encourager l'entraînement d'algorithmes fort complexes et l'estimation de très nombreux paramètres. Le besoin de stocker tous ces paramètres sur tous les n\oe uds ou calculateurs d'un nuage {\it Hadoop} impose des contraintes mémoire très fortes et donc de restreindre la taille des modèles et en conséquence la capacité d'ajustement puis de prévision par rapport à une architecture intégrée qui regrouperait mémoire et processeurs dans une seule unité. \`A ce jour et d'expérience, seule l'implémentation de la factorisation non négative de matrices ({\it NMF} pour les systèmes de recommandation) présente un réel intérêt, précision et temps de calcul, par rapport à une librairie pour l'apprentissage automatique comme {\it Scikit-learn} de Python. 

\subsection{\it Jupyter}
Le développement de codes est plus efficace dans un environnement adapté, un IDE ({\it integrated development environment}) comme par exemple {\it Eclipse}. Néanmoins pour un utilisateur et pas un développeur, cet IDE est trop complexe. Par ailleurs, l'actualité est trop souvent défrayée par des problèmes de non reproductibilité de résultats scientifiques pourtant publiés, voire même de fraudes scientifiques. Elle peuvent être dues à une simple falsification des données ou à un usage inadéquat voire malveillant de méthodes statistiques. Pour lutter, à son niveau, contre cette tendance désastreuse, le statisticien / {\it data scientist} (cf. Donoho, 2015) se doit d'écrire des codes qui rendent facile la {\it reproductibilité} de l'analyse parallèlement à la publication des données. L'environnement \href{https://rmarkdown.rstudio.com/}{\it Rmarkdown} offre de telles perspectives de même que les {\it notebooks} ou calepins de \href{http://jupyter.org/}{\it Jupyter}. Ces derniers permettent d'inclure codes, commentaires ({\it markdown}), formules en \LaTeX et résultats graphiques au sein d'un même fichier ré-exécutable pas à pas en enchaînant les clics. Ces sortes d'IDE sont tellement pratiques, ouverts à tout langage interprété (Python, R, Julia, Scala...), qu'ils deviennent le cadre de la grande majorité des ressources pédagogiques et tutoriels disponibles sur internet. C'est le choix opéré dans \href{https://github.com/wikistat/}{Wikistat 2.0}. Noter que Jupyter développe également une interface \href{https://github.com/jupyterlab/jupyterlab}{\it Jupyter lab} présentant le calepin en association à d'autres fenêtres comme dans {\it RStudio} ou {\it Matlab} et visualisant une matrice, un fichier {\tt csv.} ou une image.

\subsection{\it GitHub}
Il est important de former les étudiants à une gestion de projet {\it agile} dans un environnement de travail coopératif adéquat. L'apprentissage et l'utilisation des fonctionnalités de \href{https://git-scm.com/}{\it Git} sont donc fortement conseillées si ce n'est imposées. Le site offrant un tel service gratuitement, lorsque les dépôts sont publics, et le plus utilisé est \href{https://github.com/}{\tt GitHub}. Le succès de ce site est principalement la conséquence d'une utilisation massive par une grande majorité des développeurs de logiciels libres. Bien évidemment le rachat par {\it Microsoft} est assez antinomique et il suffirait de peu de choses, de peu de modifications dans le fonctionnement de ce site pour provoquer la fuite de tous ces développeurs / collaborateurs vers un autre site. Il suffira de suivre le mouvement mais actuellement, {\it GitHub} reste une référence et même une vitrine pour afficher des compétences de développeur ou de {\it data scientist} sur son CV.

\subsection{{\it GPU} et {\it Cloud computing}}
Les derniers choix technologiques concernent l'environnent matériel de travail. Il devient assez facile à un étudiant de s'équiper avec un ordinateur portable mais il est parfois plus fiable de disposer de salles afin d'uniformiser l'environnement de travail. R est très facile à charger et installer pour tout système d'exploitation. En revanche, les différentes versions de Python, les dépendances complexes entre les librairies, peuvent faire émerger des problèmes. Ceci se complique lorsqu'il s'agit d'utiliser des algorithmes plus sophistiqués comme {\it XGBoost} ou l'API {\it PySpark}. Dans ces derniers cas, un environnement Linux ({\it e.g. Ubuntu}) est vivement conseillé; c'est plus rarement le choix des étudiants pour leur poste personnel. Enfin entraîner des algorithmes complexes d'apprentissage profond avec {\it TensorFlow} ou optimiser finement les hyper-paramètres de {\it XGBoost} rendent indispensable l'accès à une carte graphique GPU. Certes quelques étudiants {\it gamers} en disposent mais ils sont des exceptions et le coût unitaire d'une machine devient conséquent. Pour toutes ces raisons, nous gérons deux salles de travaux pratiques au sein du département de Mathématiques dont la moitié des postes viennent d'être équipés de cartes GPU. 

Ils ne sont pas encore opérationnels mais des tutoriels ou séances de travaux pratiques vont être montés pour initier les étudiants à l'utilisation de services de {\it cloud computing}. Deux solutions sont en cours de test et d'évaluation: {\it Amazon Web Service} (AWS) et {\it Google cloud}. Le site \href{https://www.rosettahub.com/welcome}{\tt rosettaHub} offre un accès à AWS en mutualisant, entre un ensemble d'étudiants d'un même établissement, les forfaits d'utilisation gratuite mais les capacités de calcul semblent limitées. Les possibilités offertes par la \href{https://cloud.google.com/edu/}{\it Google Cloud Platform Education} sont séduisantes. Par ailleurs, l'industrie aéronautique locale, ou plutôt sa composante recherche surtout en analyse d'image, utilise majoritairement {\it Google cloud} tandis que {\it Continental Automotive} a fait le choix d'AWS pour la gestion industrielle de ses données de suivi de production. L'acquisition de compétence visée est la suivante: comment, une fois un projet protoptypé sur un poste personnel, le transférer sur un site de {\it cloud computing} pour le passage à une échelle opérationnelle; une introduction à \href{https://www.docker.com/}{\it Docker} sera envisagée dans un proche avenir afin d'automatiser au mieux le transfert.

\section{Objectifs pédagogiques}
Le facteur essentiel à prendre en compte est la très forte volatilité des méthodes, algorithmes et surtout celle des technologies concernées par l'intelligence artificielle. D'une année sur l'autre il faut être capable d'intégrer de nouveaux avatars d'algorithmes devenus incontournables dans certains domaines et les technologies afférentes. Toujours plus flexible et véloce, il faut pouvoir adapter, en cours d'année, les supports pédagogiques à la dernière version d'une librairie; l'environnement collaboratif et agile {\it Git} est indispensable à la satisfaction de ces contraintes.

{\it Attention}, ce serait une erreur de limiter la formation aux seuls aspects technologiques. Le contenu de cet article peut prêter à confusion mais il s'agit bien d'une formation d'ingénieurs de spécialité Mathématiques appliquées intégrant par ailleurs des cours fondamentaux en Optimisation, Probabilités, Statistique, Signal (Fourier, spline, ondelettes)... indispensables à une compréhension fine des algorithmes et méthodes décrites, de leurs propriétés théoriques, de leurs limites. Contenus fondamentaux qui ne se démodent pas mais difficiles à faire appréhender aux étudiants en dehors de cours et travaux dirigés finalement plus traditionnels même si des pédagogiques actives par projet ou cours inversés sont introduites.

\subsection{Former à l'autoformation}
Ce besoin de veille et d'adaptation en temps réel aux dernières méthodes et technologies restera un objectif prioritaire pour les étudiants une fois entrés dans le monde du travail. Il ne s'agit donc pas de former simplement les étudiants à des méthodes et des technologies mais plutôt de les entraîner à les apprendre par eux-mêmes avec les outils adéquats et efficaces: des tutoriels sous la forme de calepins ({\it notebooks jupyter}) sont du type de ceux que l'on trouve à profusion sur internet.

Il s'agit donc de répondre à la question: comment faire acquérir des compétences d'auto-formation aux étudiants et comment les évaluer? Une piste possible à creuser consiste à mettre les étudiants dans la situation que nous rencontrons nous-mêmes, enseignants / chercheurs, dans notre recherche ou pour simplement mettre à jour nos connaissances et nos compétences. Elle passe nécessairement par la recherche des bonnes références bibliographiques et ressources pédagogiques. La mise à disposition d'une sélection soigneuse de telles ressources sous le format de tutoriels pointant (hyperliens) vers des supports de cours ou vignettes s'avère un choix efficace car offrant un ensemble cohérent de connaissances à découvrir et mettre en \oe uvre sur un ensemble de cas d'usages réalistes afin d'acquérir les connaissances et compétences visées.

\subsection{Motivation et jeu sérieux}
Fournir des ressources pédagogiques adaptées aux étudiants est un premier pas. Leur faire utiliser ces ressources, les amener à collaborer, à auto-apprendre avec une grande part d'autonomie, à s'engager, en bref les motiver à travailler, est le deuxième pas. Cette étape, au c\oe ur du processus apprendre à apprendre, repose depuis l'année académique 2015-2016 sur un jeu sérieux basé sur un concours de prévision par apprentissage automatique. 

Cette notion de problème à résoudre dans le contexte d'un concours n'est pas originale, elle a été largement popularisée par le concours {\it Netflix} de recommandation de films avec un prix d'un million de dollars mais pré-existait dans d'autres contextes: concours d'analyse et de description de données par l'Association Américaine de Statistique, concours de prévision lors des congrès de chimiométrie. Actuellement le site {\it Kaggle} (racheté par {\it Google} en mars 2017) affiche régulièrement des offres de concours primés ou non. Cette idée a été reprise en France à des fins pédagogiques par l'\href{https://challengedata.ens.fr/en/home}{École Normale Supérieure}, l'\href{https://www.datascience.net/fr/home/}{ENSAE ParisTech} ou l'\href{http://www.datascience-paris-saclay.fr/new-ramp-imaging-psychiatry-challenge-impac/}{Université Paris-Saclay}. Nous testons ou faisons tester par les étudiants (projets tutorés) ces concours depuis leur création afin d'en apprécier les intérêts pédagogiques et limites mais la mise en place d'un concours ou défi spécifique fut la réponse adaptée à notre environnement et nos objectifs. 

En effet, plusieurs raisons: 
\begin{itemize}
\item maîtrise du sujet, de son niveau de complexité, 
\item maîtrise du calendrier de début (octobre) et fin (janvier) de concours, 
\item émulation nettement plus forte entre pairs plutôt qu'en concurrence avec des professionnels anonymes, 
\end{itemize}
nous ont poussés à l'organisation locale d'un défi toulousain, rejoint successivement ensuite par des équipes de masters des universités de Bordeaux, Pau puis Rennes, Nantes, Paris, Lyon. Ce sont plus de 40 équipes de 4 à 5 étudiant-e-s qui se sont affrontées entre octobre 2016 et mi-janvier 2017, 55 pour le {\it défi grosses data} de 2018. D'autres masters s'associant à ce projet, nous attendons la participation de plus de 300 étudiants pour le \href{https://defi-ia.insa-toulouse.fr/}{\it défi IA 2019}.

Les sujets et jeux de données afférents évoluent chaque année:
\begin{itemize}
\item 2016: construction d'un système de recommandation de films à partir des données publiques du site {\it Movie Lens},
\item \href{https://maxhalford.github.io/blog/a-short-introduction-and-conclusion-to-the-openbikes-2016-challenge/}{2017}: prévision du nombre de vélos sur un ensemble de stations de vélos en libre service dans différentes villes: Paris, Lyon, Toulouse puis New-York,
\item \href{https://defibigdata2018.insa-toulouse.fr}{2018}: Collaboration avec Météo France pour la prévision de températures par adaptation statistique des modèles déterministes ARPEGE et AROME,
\item \href{https://defi-ia.insa-toulouse.fr/}{2019}: Collaboration avec AIBUS DS pour détecter la présence ou non d'une éolienne sur une photo satellite.
\end{itemize}

Comme sur les sites de type {\it Kaggle}, le concours est organisé en deux phases. Dans la première (3 mois), le classement est public, établi sur une partie de l'échantillon test, et chaque équipe tente de progresser dans le classement régulièrement mis à jour en testant, expérimentant, les méthodes accessibles et décrites dans les tutoriels. La deuxième étape est le classement final, fourni par la dernière solution de chaque équipe appliquée à la partie confidentielle de l'échantillon test de façon à traquer un possible sur-apprentissage, piège bien connu de ces techniques à faire expérimenter par les étudiants.

La motivation, l'engagement et l'entrain des étudiants ont été au rendez-vous comme le montre régulièrement la restitution des résultats lors d'une journée associant exposés académiques, industriels et présentations des solutions les plus performantes. 

\subsection{Évaluation}
L'évaluation de l'unité de formation associée au défi découle facilement de son organisation. Les performances des solutions proposées assorties d'un exposé oral permettent de valider finement quelles sont les compétences acquises. Bien entendu tous les étudiants ne sont pas des {\it geeks} passionnés consacrant leurs soirées à améliorer leur solution pour grignoter des places dans le classement. Pour celles et ceux, moins motivés par l'esprit de compétition d'un concours, l'évaluation de l'UF repose toujours sur la présentation orale de la démarche mise en place et aussi sur un objectif de résultat ({\it base line}) à atteindre {\it a minima}.  L'objectif de ce résultat garantit en effet que le groupe d'étudiants maîtrise les bases indispensables et nécessaires à la mise en \oe uvre des algorithmes d'apprentissage et à leur optimisation; la présentation orale témoigne qu'ils sont capables d'en rendre compte, qu'ils maîtrisent leur sujet en expliquant les options et choix qu'ils ont été amenés à engager.

\section{Contenus et ressources pédagogiques}
\vspace*{1mm}
\centerline{\includegraphics[width=4cm]{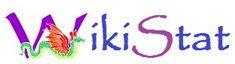}}
\vspace*{1mm}

Les projets de contenus pédagogiques de la spécialité Mathématiques Appliquées de l'INSA de Toulouse ont déjà été introduits par Besse et Laurent (2015); ils sont maintenant en place avec évidemment quelques adaptations. Les contenus détaillés (fiches ECTS) sont disponibles sur le \href{http://www.insa-toulouse.fr/fr/formation/ingenieur/offre-de-formation-ects/sciences-technologies-sante-STS/formation-d-ingenieur-FI/ingenieur-specialite-genie-mathematique-et-modelisation-program-fruai0310152xpri5mm000.html}{site de l'Établissement}. Nous ne détaillerons pas toutes les Unités de Formations (UFs) mais donnerons quelques indications sur l'organisation des plus spécifiques à l'enseignement de l'IA et prenant en compte  l'environnement industriel toulousain, source de très nombreuses offres de stages et d'emplois, principalement depuis 2017.

\subsection{Unités de formation en IA}
Historiquement, l'ancienne spécialité Génie Mathématique et Modélisation, devenue Ma\-thématiques Appliquées en 2018, avait deux orientations, l'une déterministe ou numérique, l'autre statistique et stochastique. Ce clivage est hérité de la structuration des équipes académiques de recherche toulousaines mais c'est un anachronisme pour le monde industriel. Combler ce fossé, lors de la mise en place de la dernière maquette a conduit à renforcer le tronc commun de 4ème année (M1), notamment les enseignements d'optimisation, analyse fonctionnelle, traitement du signal, images. Avec la même finalité, des UFs sont optionnelles en 5ème année (M2) mais celle basique d'{\it apprentissage automatique} est choisie par tous les étudiants quelque soit leur majeure. 

Ces choix sont fortement déterminés par le tropisme aéronautique toulousain. Jusqu'en 2015, 2016, la majorité des emplois de {\it data mining, data science} étaient proposés dans la région parisienne, principalement en lien avec le marketing et la vente en ligne. Depuis la promotion 2017, nous assistons à une très forte demande locale dans le secteur industriel, aéronautique et sous-traitance, mais aussi automobile (Continental Automotive, Renault software lab). Cette demande a largement conforté les choix du comité d'orientation du département pour renforcer les thématiques pour les applications industrielles plutôt que la vente en ligne. Cet objectif a convergé naturellement avec le rapprochement des deux orientations, déterministes et stochastiques du département pour apporter des réponses concrètes aux besoins industriels actuels bien identifiés: analyse de sensibilité, méta-modèles, détection d'anomalies, maintenance prédictive, qui sont de plus assortis de structures de données  de très grande dimension: signaux, courbes, images. Ces spécificités locales ont largement motivé les renforcements méthodologiques de la 4ème année (optimisation non différentielle, optimisation stochastique et séquentielle, signal, image) de même que ceux technologiques de la 5ème.

En plus des UFs avec des contenus spécifiques dont certaines optionnelles: plan d'expérience, fiabilité, analyse de sensibilité, image, calcul stochastique, assimilation de données, trois UFs de 5ème année (M2) concernent plus particulièrement les applications de l'IA. Elles sont structurées en trois couches. 
\subsubsection{Apprentissage automatique ou statistique}
La première couche est suivie par tous les étudiants, c'est l'UF de base qui décrit les propriétés de l'ensemble des méthodes et algorithmes d'apprentissage les plus utilisés, de la régression logistique aux forêts aléatoires en passant par l'analyse discriminante et les $k$ plus proches voisins, les arbres binaires de décision, le {\it boosting}. Une introduction est proposée aux supports à vaste marge (SVM) et aux réseaux de neurones; les méthodes d'imputation de données manquantes ainsi que celles de détection d'anomalies sont également traitées de même que les questions éthiques: biais et discrimination, droit à l'explication, mises en exergue par le déploiement du RGPD (règlement général sur la protection des données).

\subsubsection{Apprentissage en grande dimension}
La deuxième couche est suivie par les étudiants de l'orientation stochastique. Elle vise à compléter la précédente en apportant des éléments théoriques et méthodologiques spécifi\-ques aux données de grande dimension: signaux et images. Ce cours aborde les notions de sélection de modèles et sélection de variables dans un modèle linéaire en grande dimension (pénalisation Ridge, Lasso...), les méthodes de classification linéaire et non linéaire, en particulier les SVM. L'agrégation de classifieurs est également introduite. Nous abordons les méthodes classiques en régression non paramétrique : régressogramme,  estimateurs à noyau, splines, estimateurs par projection sur des bases orthonormées (Fourier, ondelettes), estimation par seuillage sur des bases d'ondelettes, ceci afin de traiter des données fonctionnelles ou des images. Enfin, les réseaux de neurones et une introduction au {\it deep learning}, notamment aux réseaux convolutionnels sont traités. Les cours sont associés à un volume équivalent de travaux pratiques sous forme de tutoriels en Python ou R. La détection d'anomalies dans des données fonctionnelles est également traitée lors des travaux pratiques. 

En plus d'un contrôle rapide sur table, ces deux UFs sont évaluées par le biais d'un projet réalisé en deux étapes par les étudiants, une par UF.   La nature ou plutôt la complexité des données qui sont renouvelées chaque année justifie de cette organisation. Le sujet 2017-2018 concernait l'analyse de données (accéléromètre, gyroscope) issues d'un smartphone pour identifier l'activité de son porteur. Il s'agissait d'un problème de classification supervisée à traiter sur la base de données transformées à l'aide d'algorithmes d'apprentissage automatique usuels pour la première UF tandis que l'analyse des données brutes à l'aide de neurones profonds concernait la deuxième UF. Ce cas d'usage est maintenant un tutoriel qui sert de fil rouge pour les travaux pratiques ainsi que pour des actions de formation continue. Le sujet 2018--2019 portera également sur l'analyse  de signaux mais cette fois physiologiques (EEG) pour de l'aide au diagnostic. 

\subsubsection{Technologies de l'IA}
La troisième couche est spécifique aux étudiants de la majeure {\it Science des Données}. Elle apporte les compétences nécessaires à l'utilisation des outils technologiques récents de traitement des données massives ({\it Spark}) mais aussi de l'apprentissage profond ({\it Keras, TensorFlow}) ou du {\it cloud computing} ({\it Google Cloud}). L'étudiant est amené à développer une démarche critique quant au choix ou non d'utiliser ces technologies, ce qui n'est pas toujours nécessaire.  Elle vise également à rapprocher le plus possible l'étudiant de l'expérience pratique du métier de {\it data scientist} et du travail de manipulation et de mise en forme des données ({\it data munging}) qui précède l'application d'algorithmes d'apprentissage.

Elle est basée sur l'analyse de cas d'usage qui abordent des types de données différents et complémentaires sur lesquels des algorithmes récents ont déjà fait leur preuve. La classification d'images ({\tt CatsVsDogs}) et la reconnaissance de caractères ({\tt MNIST}), à l'aide de réseaux de neurones convolutifs y sont traités. L'étudiant est également amené à manipuler des réseaux pré-entraînés: {\tt ResNet, Inception} (transfert d'apprentissage). Nous abordons également le traitement automatique du langage naturel (NLP) à travers la classification supervisée de descriptions textuelles de produits (Cdiscount) en présentant les différentes méthodes de pré-traitement du texte ({\it Racinisation, Tokenizing}), la vectorisation de ces données ({\it term frequency minus inverse document frequency, Word embedding, Word2vec}), et l'application d'algorithmes d'apprentissage sur ces données. Les réseaux récurrents (LSTM) sont  appliqués sur ces données pour la génération automatique de contenu. La recommandation de Films ({\tt movieLens}) par filtrage collaboratif à l'aide de méthodes de factorisation (décomposition en valeurs singulières, factorisation non négative de matrices) est également étudiée.

\subsection{\href{http://wikistat.fr/}{Wikistat 1.0}}
\begin{figure}
\centerline{\includegraphics[width=12cm]{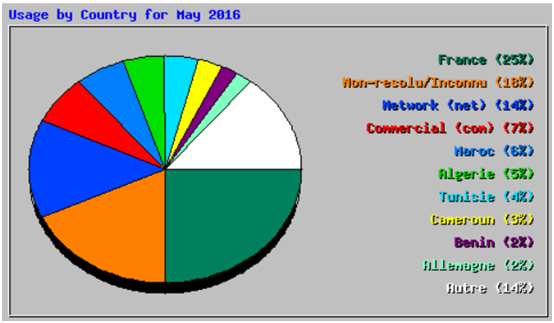}}
\caption{\it Répartition géographique des consultations de {\tt wikistat.fr} montrant une présence significative de l'Afrique francophone.}\label{pieWik}
\end{figure}

Les ressources pédagogiques de ces trois UFs sont disponibles en ligne comme celles de l'UF de 4ème année: logiciels et exploration statistiques. Un premier site: \href{http://wikistat.fr/}{\tt wikistat.fr} regroupe depuis plusieurs années des ressources mises à disposition par les intervenants. Elles prennent la forme de vignettes (fichier pdf issu d'un source \LaTeX) de cours et travaux pratiques; vignettes par méthode, algorithme ou famille de méthodes. Ce site, sert et servira toujours de référence de cours. Il est largement consulté par les étudiants locaux ainsi que ceux francophones comme le montrent quelques statistiques du tableau \ref{tabWik} et de la figure \ref{pieWik} relatives au mois de mai 2016. Il est actuellement toujours aussi consulté mais l'ouverture du deuxième site en biaise les statistiques.

\begin{table}
\caption{\it Nombre de consultations des 5 documents les plus chargés en mai 2016 sur le site hébergeant physiquement {\tt wikistat.fr}. Les deux premiers sont des polycopiés obtenus par simple compilation des vignettes de wikistat.fr.}\label{tabWik}
\centerline{\includegraphics[width=12cm]{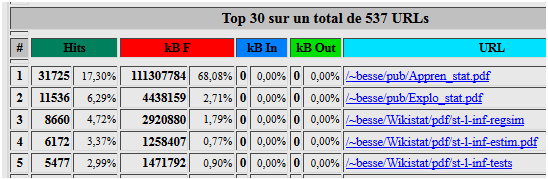}}
\end{table}

Les supports pédagogiques en anglais sont légion sur le web mais nettement moins accessibles en français et au niveau M2 visé; ce qui explique le succès de ce site.

\subsection{\href{http://wikistat/}{Wikistat 2.0}}
Le site précédent {\tt wikistat.fr} est déjà ancien et bien référencé par les moteurs de recherche mais il nécessitait une refonte pour être  intégré à un environnement collaboratif professionnel. Le choix stratégique, économique en temps de travail, a été de l'intégrer ou plutôt d'intégrer virtuellement tous les supports de cours (vignettes) existants comme  cibles de liens hypertextes présents dans les tutoriels (calepins ou {\it jupyter notebooks}) ou cas d'usage du site {\tt github.com/wikistat}. L'ensemble des ressources pédagogiques est donc accessible selon deux entrées: à partir d'un exposé classique de cours en présentiel et y faisant référence ou à partir de cas d'usage exécutables en présentiel (travaux pratiques) ou en autonomie (tutoriels). La mise en place du site a bénéficié d'une aide à l'innovation pédagogique de l'INSA de Toulouse pour être rendu rapidement opérationnel.

Le site est structuré en 5 saisons découpées en épisodes déroulant la chronologie classique d'un cours (ficher {\tt README}) à travers des tutoriels (calepins) pouvant être utilisés en autoformation ou comme support de travaux pratiques. Les UFs précédentes constituent les saisons 3 à 5. Le format des calepins ({\it notebook}) permet d'intégrer, selon les besoins ou nécessités pédagogiques, des exercices ou de simples questions afin de motiver la réflexion des impétrants au delà d'une simple exécution. 

\section{Conclusion}
Les retours des étudiants à propos de ces ressources de cours, des tutoriels et plus particulièrement du défi sont dans l'ensemble très positifs avec des remarques prises en compte chaque année afin d'améliorer le dispositif. Ainsi, les étudiants, au moins ceux toulousains, se rencontrent début octobre lors du lancement du défi et lors de la séance de restitution. Les présentations orales de l'organisation générale, des données et des objectifs sont filmées et rendues  accessibles aux participants des autres universités. Le site affiche une représentation graphique de l'évolution des classements publics. Les calendriers des différents établissements sont coordonnés au mieux des possibilités et flexibilités de chacun.

Le principal objectif de motivation des étudiants par le défi est atteint; celle-ci a largement dépassé le seul but trivial de validation d'une unité de formation. Ce n'est pas non plus un classement de sortie ancestral qui pousse les étudiants à travailler de façon solitaire. C'est plutôt la réussite d'un projet conduit par une équipe de façon agile et collaborative, pimentée par la compétition entre formations thématiquement voisines.

Cette motivation permet d'atteindre l'objectif d'auto-apprentissage. Les solutions expé\-rimentées et rendues opérationnelles ont largement exploité les programmes prévus des UFs d'apprentissage statistique. Les étudiants ont spontanément approfondi des notions (optimisation, pénalisation) ou des technologies (Python, PySpark, librairies R, GPU, {\it cloud computing}) qui se sont avérées ou s'avéreront indispensables au bon déroulement des projets car ils sont directement confrontés aux problèmes posés par un passage aux échelles volume et variété des données.

Le projet pédagogique relaté dans cet article se construit en avançant, expérimentant. Les retours des étudiants sont très positifs, leur insertion actuelle, dans les très nombreux stages faisant appel aux compétences visées, est très bonne. L'insertion professionnelle l'est également. Ainsi, pour la spécialité Mathématiques Appliquées de l'INSA de Toulouse et pour les promotions 2016 et 2017, 75 \% avaient signé un contrat avant la fin de leur stage, 50 \% ont un CDI. Tous les retours nous incitent à poursuivre et développer cette expérience. 

Attention, il faut rester prudent et réaliste, les ressources pédagogiques mises à disposition requièrent une veille technologique permanente, et donc des moyens humains afférents, pour résister à une obsolescence très rapide des technologies mises en \oe uvre. C'est la condition incontournable pour préserver l'excellente insertion actuelle des étudiants en anticipant de possibles et très probables retournements de conjonctures. Des bases mathématiques solides et largement immuables ainsi que des compétences technologiques de pointe sont les deux piliers de notre stratégie pour continuer à surfer, ou faire surfer par les étudiants, la vague des données massives en exploitant les méthodes de l'IA.

\vspace*{5mm}
\centerline{\includegraphics[width=8cm]{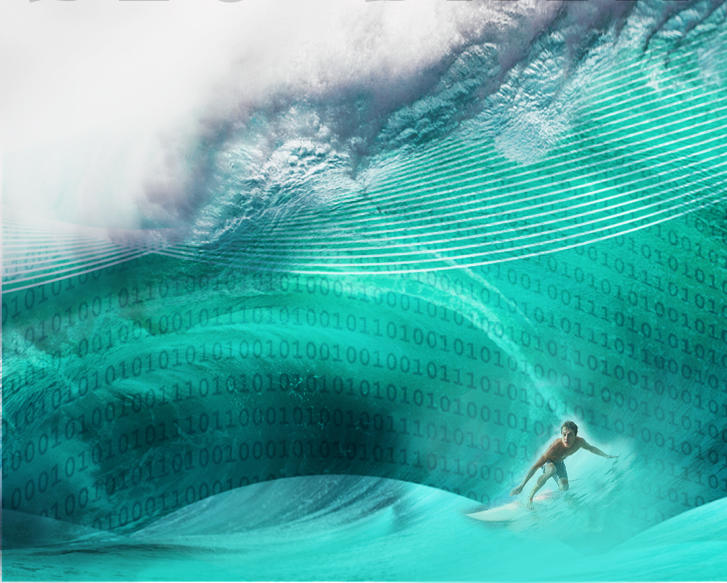}}

\end{document}